\documentclass[11pt,preprintnumbers,pra,aps,
nofootinbib,tightenlines,floatfix, twocolumn,superscriptaddress,longbibliography]{revtex4-2}

\usepackage[colorlinks=true]{hyperref}
\hypersetup{
  linkcolor  = TealBlue!90!Black,
  citecolor  = YellowGreen!90!Black,
  urlcolor   = Peach!90!Black,
  colorlinks = true,
}
\usepackage[dvipsnames,table]{xcolor}
\usepackage[all]{hypcap}
\usepackage{amsmath,amssymb}
\usepackage{graphicx}
\usepackage{balance}
\usepackage[utf8]{inputenc}
\usepackage{csquotes}
\usepackage{empheq}
\usepackage{multirow}
\usepackage[section]{placeins}
\usepackage{comment}
\usepackage[customcolors]{hf-tikz}
\usepackage[top=1in,bottom=1in,left=1.8cm,right=1.6cm]{geometry}

\allowdisplaybreaks
\makeatletter
\renewcommand\onecolumngrid{
\do@columngrid{one}{\@ne}
\def\set@footnotewidth{\onecolumngrid}
\def\footnoterule{\kern-6pt\hrule width 1.5in\kern6pt}%
}
\makeatother

\newcommand\scalemath[2]{\scalebox{#1}{\mbox{\ensuremath{\displaystyle #2}}}}

\newcount\hour \newcount\hourminute \newcount\minute
\hour=\time \divide \hour by 60
\hourminute=\hour \multiply \hourminute by 60
\minute=\time \advance \minute by -\hourminute

\begin{document}

\title{Identification of a natural fieldlike entanglement resource in trapped-ion chains}
\author{Natalie Klco}
\email{natalie.klco@duke.edu}
\affiliation{{Duke Quantum Center and Department of Physics, Duke University, Durham, NC 27708, USA}}
\author{D.~H.~Beck}
\email{dhbeck@illinois.edu}
\affiliation{Department of Physics and Illinois Quantum Information Science and Technology (IQUIST) Center, University of Illinois at Urbana-Champaign,  Urbana, IL 61801, USA}

\preprint{IQuS{\_}WKSHP@UW-23-001}

\begin{abstract}
The electromagnetic trapping of ion chains can be  regarded as a process of non-trivial entangled quantum state preparation within Hilbert spaces of the local axial motional modes.
To begin uncovering properties of this entanglement resource produced as a byproduct of conventional ion-trap quantum information processing, the quantum continuous-variable formalism is herein utilized to focus on the leading-order entangled ground state of local motional modes in the presence of a quadratic trapping potential.
The decay of entanglement between disjoint subsets of local modes is found to exhibit features of entanglement structure and responses to partial measurement reminiscent of the free massless scalar field vacuum.
With significant fidelities between the two, even for large system sizes, a framework is established for initializing quantum field simulations via \enquote{imaging} extended entangled states from natural sources, rather than building correlations through deep circuits of few-body entangling operators.
By calculating probabilities in discrete Fock subspaces of the local motional modes, considerations are presented for locally transferring these pre-distributed entanglement resources to the qudits of ion internal energy levels, improving this procedure's anticipated experimental viability.
\\
\\
\end{abstract}
\maketitle

{
\tableofcontents
}

\begin{quote}
\begin{center}
\enquote{\emph{In endeavors of discovery \hspace{1.5cm} \ \\it is wise to listen  \hspace{1cm} \ \\  as Nature speaks for herself.}}
\end{center}
\end{quote}

\section{Introduction}
\label{sec:introduction}

Spatially distributed quantum correlations not only play an important role in the structure, observed dynamics, and apparent thermalization of quantum many-body systems~\cite{Ho:2015rga,2016Sci...353..794K,Kharzeev:2017qzs,Baker:2017wtt,Cervera-Lierta:2017tdt,Berges:2018cny,Beane:2018oxh,Johnson:2022mzk,Beane:2019loz,Tu:2019ouv,Beane:2020wjl,Beane:2021zvo,Kharzeev:2021yyf,Robin:2020aeh,Low:2021ufv,Gong:2021bcp,Roggero:2021asb,Mueller:2021gxd,Joshi:2023rvd}, but act as a fundamental resource in designing and operating quantum computation, simulation, sensing, and communication technologies~\cite{Feynman1982,Plenio:2007zz,Nielsen2010-ga,Horodecki2009QE,wilde2013quantum,2017RvMP...89c5002D,Klco:2021lap,Kaubruegger2021ram,Marciniak:2021tld,PRXQuantum.2.017003,Daley2022,Brady:2022bus}.
For such applications, the necessary entanglement is commonly produced via a series of direct interactions generated between two or more quantum degrees of freedom, or distributed after production through quantum networking techniques that maintain coherence throughout transmission.
\begin{figure}
    \centering
    \includegraphics[width=0.45\textwidth]{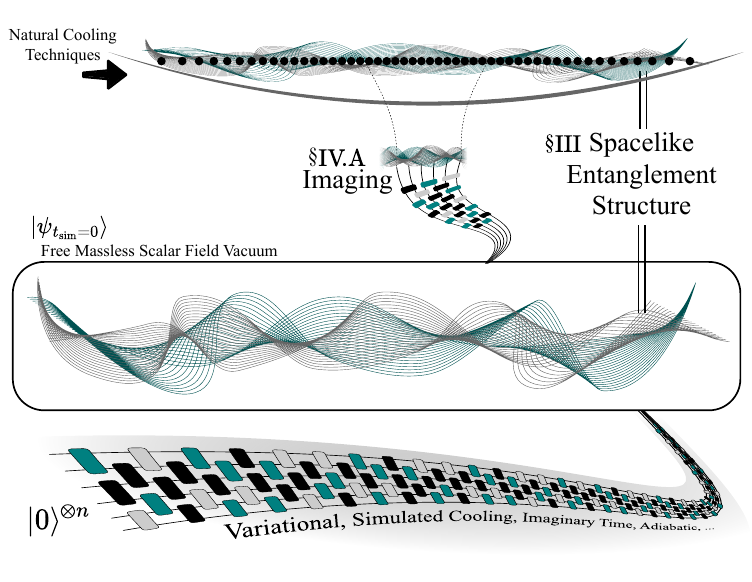}
    \caption{
    (top right) As discussed in Section~\ref{sec:entanglementresource}, the structure of spacelike distributed entanglement between disjoint regions of the free massless scalar field vacuum is similar to that of the axial modes in a trapped ion chain motional ground state.  
    (top center) Motivated by this observation, the cooled motional modes are proposed in Section~\ref{sec:prelimapps} as an entanglement resource that may be 
    \enquote{imaged} as an alternative method of vacuum state preparation for scattering and quantum field simulations.}
    \label{fig:fig0}
\end{figure}
A third possibility, however, relies upon connecting locally to a shared system of naturally distributed entanglement, see Fig.~\ref{fig:fig0}.
For example, the vacuum state of a quantum field is entangled at spacelike separations~\cite{Reeh1961,summers1985vacuum,summers1987bell1,summers1987bell2,Halvorson:1999pz}, and a pair of locally-interacting detectors can transfer that entanglement into basic quantum degrees of freedom~\cite{VALENTINI1991321,Reznik:2002fz,Reznik:2003mnx} to drive subsequent quantum information processing protocols.
While the raw entanglement production through this type of extraction is expected to be far smaller than can be generated through artificially constructed logic gates, \enquote{imaging} natural entanglement can provide a valuable structure of quantum correlations, e.g., as a starting point for large-scale quantum simulations of physical systems or the generation of entangled sensing arrays.
The present paper focuses on the local axial motional modes of a trapped ion chain, quantifying their entanglement properties and utilizing this information to guide applications in the quantum simulation of quantum fields.

With each global motional mode in its ground state, the vacuum of ion-chain zero-point fluctuations is an unentangled, tensor-product state with respect to partitions among the normal modes.
However, for partitions among the associated local motional modes, this vacuum exhibits entanglement and, to a good approximation for practical trapping potentials, constitutes a Gaussian continuous-variable (CV) system.
Note that the ion chain supports both transverse and axial modes (see Ref~\cite{Chen:2021ChPhB} for a recent review).
While the global motional modes are commonly utilized to mediate all-to-all entangling gates among ions within the trap~\cite{Cirac:1995zz,Monroe:1995dem,Sorensen1999thm,Leibfried:2003exp,SchmidtKaler:2003real,Katz:2022ncq}, the local transverse motional modes have been proposed to represent computational bosonic degrees of freedom in the design of lattice gauge theory quantum simulations~\cite{Casanova:2011wh,Davoudi:2021ney}.

For a pair of disjoint regions within the global vacuum state, each comprising the fluctuations associated with one or more consecutive ions, it is herein found that many entanglement characteristics are remarkably similar to those of the massless non-interacting lattice scalar field vacuum.
In particular, the logarithmic negativity between the regions decreases exponentially with increasing separation, whereas the corresponding two-point functions decrease, at most, as a power law.
As with the scalar field vacuum, the stronger entanglement represented by the two-point functions can also be recovered with classical communication of measurements on parts of the system external to the two regions~\cite{Klco:2023ojt}.

After presenting the local axial modes of the trapped-ion vacuum as a Gaussian CV system in Section~\ref{sec:formalism}, its entanglement characteristics are described with reference to those of the scalar field vacuum in Section~\ref{sec:entanglementresource}. Two possible applications are subsequently discussed in Section~\ref{sec:prelimapps}, using the trapped ion vacuum both as a state preparation resource for the scalar field vacuum and as a more general resource from which entanglement can be transferred into the discrete-variable framework of ion internal states~\footnote{Note that the scalar field is also referred to as a harmonic chain or Klein-Gordon field.}.

In pursuing these applications, 
entanglement structure provided essential guidance, e.g., by indicating that the scalar field vacuum and axial modes of the ion chain motional ground state naturally contain similar distributions of quantum correlations, or by identifying the dominant collective modes to optimize entanglement transfer.
While the connections explored here represent only early observations, this concrete example suggests significant opportunities for incorporating natural sources of distributed entanglement as active components in quantum technologies.

\section{Local Axial Motional Modes}
\label{sec:formalism}

The Hamiltonian describing $N$ ions in a 1D\nobreakdash-linear trap is the sum of the ion kinetic energies~($T$) and Coulomb interactions within the trapping potential~($U$),
\begin{align}
T &= \sum_{i=1}^N \frac{m\dot{z}_i^2}{2}  \\
U &=  \sum_{i = 1}^N q \kappa_2 z_i^2 + \frac{1}{2} \sum_{\stackrel{i,j=1}{i\neq j}}^N \frac{q^2 }{4 \pi \epsilon_0 |z_i - z_j|}   \ \ \ ,
\end{align}
where $q, m$ are the charge and mass of each ion (assuming homogeneous ion species), $\kappa_2$ is the strength of the quadratic trapping potential, and $z_i$ is the position of the $i^{\text{th}}$ ion along the one-dimensional trap axis~\cite{,Raizen1992,Wineland:1997mg}.
Although the incorporation of anharmonicities in the trapping potential~\cite{Lin:2009anharmonic,Home2011:rdq} will be of future interest, e.g., for opportunities to design ion spacings that produce local motional modes with a regularity that mirrors conventional latticizations of field theories, the current work utilizes a quadratic trapping potential.

As usual, the equilibrium positions of ions within the trap may be calculated by solving the set of equations enforcing  vanishing first derivatives of the potential with respect to displacements,
\begin{equation}
\frac{\partial U}{\partial z_i} \Big\rvert_{\mathbf{z} = \mathbf{z}_0} = \vec{0} \ \ \ ,
\end{equation}
to place each ion in a stable local minimum.
For small systems, these equations can be solved exactly (see Appendix~\ref{app:smallexamples}).
For larger systems, the equilibrium positions can be reliably calculated numerically~\cite{Steane:1996ew,James:1997mn}.
The central ions of these large chains develop an approximately evenly spaced distribution, with the edges systematically relaxing to larger separations as the soft harmonic potential is countered by the center-dominated Coulomb repulsion.

The above Hamiltonian has two significant scales, one associated with the inter-ion spacing and one associated with oscillations of the ions about their equilibrium positions.
Isolating these characteristic scales to identify a dimensionless description of the system, the Hamiltonian may be written as
\begin{align}
H &= \frac{m \omega_z^2 \ell^2_{\mu m}}{2}  \sum_{i=1}^N  \left( \dot{\bar{z}}_i^2 +  \bar{z}_i^2 + \sum_{\stackrel{j=1}{j\neq i}}^N \frac{1}{|\bar{z}_i - \bar{z}_j|} \right) \\
&= \frac{m \omega_z^2 \ell^2_{\mu m}}{2} \left( \bar{T} + \bar{U} \right) \ \ \ ,
\label{eq:Hsep}
\end{align}
where $\bar{z} = \frac{z}{\ell_{\mu m}}$ and $\dot{\bar{z}} = \frac{\dot{z}}{\omega_z\ell_{\mu m}}$.  The characteristic spacing and oscillation parameters are $\ell_{\mu m}~=~\left(\frac{q}{8\pi \epsilon_0 \kappa_2}\right)^{\frac{1}{3}}$ and $\omega_z^2 = \frac{2q \kappa_2}{m}$, where the latter is the center-of-mass oscillation frequency for an ion chain.
The root-mean-square of fluctuations from equilibrium in the ground state of an oscillator of this frequency yields the length, $\ell_{nm} =  \sqrt{\frac{\hbar}{m \omega_z}}$.
To provide a sense of scale for these two lengths, a single $^{171}$Yb$^+$ ion in a trap with $\omega_z/2\pi = 1$~MHz yields fluctuations characterized by the scale $\ell_{nm}=7.7$~nm, while the separation between such ions is governed by the larger scale, $\ell_{\mu m} = 2.7~\mu$m.
As such, the lengths have been labeled for their canonical magnitudes in a broad regime of trapped ion applications.

With a second set of variables defined and scaled for excursions from equilibrium positions, $\bar{\boldsymbol{\zeta}} = \frac{\mathbf{z}-\mathbf{z}_0}{\ell_{nm}} = (\bar{\mathbf{z}} - \bar{\mathbf{z}}_0)\frac{\ell_{\mu m}}{\ell_{nm}}$, the potential may be expanded about the equilibrium as
\begin{multline}
U = \frac{\hbar \omega_z}{2} \frac{\ell_{\mu m}^2}{\ell_{nm}^2} \left(\frac{1}{2!}\frac{\partial^2 \bar{U}}{\partial \bar{z}_i \partial \bar{z}_j}\Big\rvert_{\bar{\mathbf{z}} = \bar{\mathbf{z}}_0} \left(\frac{\ell_{nm}}{\ell_{\mu m}}\right)^2 \bar{\zeta}_ i \bar{\zeta}_j \ + \right.
\\
\left.\frac{1}{3!}\frac{\partial^3 \bar{U}}{\partial \bar{z}_i \partial \bar{z}_j \partial \bar{z}_k}\Big\rvert_{\bar{\mathbf{z}} = \bar{\mathbf{z}}_0} \left(\frac{\ell_{nm}}{\ell_{\mu m}}\right)^3 \bar{\zeta}_ i \bar{\zeta}_j\bar{\zeta}_k + \cdots \right) \ \ ,
\label{eq:harmonicapprox}
\end{multline}
where a shift has been introduced to place the equilibrium configuration at zero potential, $\bar{U}\left(\bar{\mathbf{z}}_0\right)=0$.
With higher order contributions suppressed by powers of $\frac{\ell_{nm}}{\ell_{\mu m}}$, the convergence of this expansion is commonly rapid.  For the representative $^{171}$Yb$^+$ example considered above, this ratio is $2.8 \times 10^{-3}$.
Implementing a harmonic approximation of the potential, the leading-order dimensionless Hamiltonian describing small oscillations about equilibrium is
\begin{equation}
\bar{H}_2 =  \sum_{i=1}^N \dot{\bar{\zeta}}^2_i +  \sum_{i,j=1}^N \bar{\zeta}_i \bar{L}_{ij} \bar{\zeta}_j   \ \ \ ,
\label{eq:Hbar2}
\end{equation}
where $H_2 = \frac{\hbar \omega_z}{2} \bar{H}_2$ and $\bar{L}_{ij} = \frac{1}{2} \frac{\partial^2 \bar{U}}{\partial \bar{z}_i \partial \bar{z}_j}\rvert_{\bar{\mathbf{z}} = \bar{\mathbf{z}}_0}$ is the Hessian matrix of second derivatives.
In the following, this harmonicity supports the use of the Gaussian continuous variable formalism~\cite{Braunstein:2005zz,Serafini2017}, allowing the leading-order entanglement information of the ground state to be succinctly captured by the covariance matrix alone.

By diagonalizing the potential, axial normal modes of motion may be calculated as position space eigenvectors, $\mathbf{e}_\alpha$, of the Hessian with eigenvalues, $\bar{\omega}_\alpha^2$, in units of the center-of-mass oscillation frequency, ${\bar{\boldsymbol{\omega}}=\boldsymbol{\omega}/\omega_z}$.
In this basis of normal modes, the dimensionless Hamiltonian operator is that of a set of independent oscillators,
\begin{equation}
\hat{\bar{H}}_2 = \sum_\alpha \hat{\bar{\nu}}_\alpha^2  +  \bar{\omega}_\alpha^2 \hat{\bar{\xi}}_\alpha^2 \ \ \ ,
\end{equation}
with $\hat{\bar{\xi}}_\alpha$ and $\hat{\bar{\nu}}_\alpha$ the dimensionless position and momentum operators, respectively, of the chain's normal modes.
As usual, these operators may be
written in terms of normal mode creation and annihilation operators,
\begin{align}
\hat{\bar{\xi}}_{\alpha} &= \frac{\hat{\xi}_\alpha}{\ell_{nm}} = \frac{\hat{a}_\alpha + \hat{a}_{\alpha}^\dagger}{\sqrt{2\bar{\omega}_\alpha}}
\\
\hat{\bar{\nu}}_\alpha &= \frac{\ell_{nm}}{\hbar}\hat{\nu}_\alpha = -i\sqrt{\frac{\bar{\omega}_\alpha}{2}} \left(\hat{a}_\alpha - \hat{a}^\dagger_\alpha\right) \ \ \ ,
\end{align}
with $\left[ \hat{a}_\alpha, \hat{a}^\dagger_\beta\right] = \delta_{\alpha\beta}$.
Naturally, the matrix of normal mode eigenvectors, $\mathbf{e}$, constructed in rows as $ e_{\alpha,i} = \left(e_{\alpha}\right)_i$,  provides the unitary basis transformation between global and local motional modes
\begin{equation}
 \hat{\boldsymbol{\phi}} = \mathbf{e}^{-1} \hat{\boldsymbol{\xi}}
\quad , \quad
  \hat{\boldsymbol{\pi}} = \mathbf{e}^{-1} \hat{\boldsymbol{\nu}} \ \ \ ,
  \end{equation}
where $\hat{\boldsymbol{\phi}}$ and $\hat{\boldsymbol{\pi}}$ are the vectors of position and conjugate-momentum operators for the local motional modes, respectively.
\begin{figure}
    \centering
\includegraphics[width=0.45\textwidth]{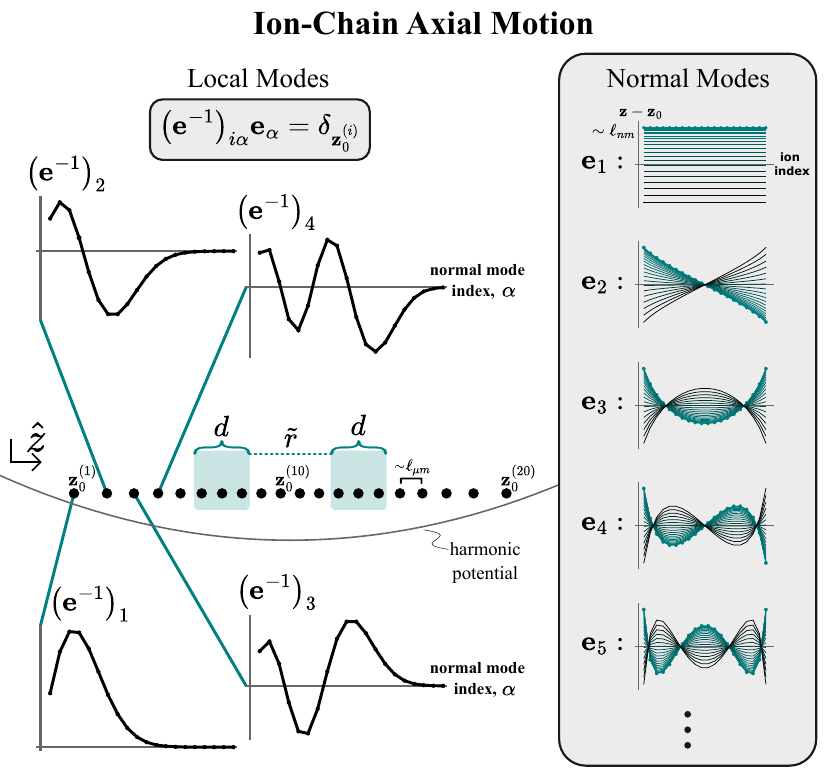}
    \caption{For ions in a linear trap (center) with equilibrium positions $\mathbf{z}_0^{(i)}$, normal modes of axial motion $\mathbf{e}_\alpha$ (right) may be linearly combined to form an alternate basis (left) describing the local motion of individual ions.  
    The center diagram shows an example configuration of symmetrically distributed regions of size $d = 3$ and separation $\tilde{r} = 4$ in a chain of $N = 20$ ions. 
    Analogous to the tensor product(entangled) structure of momentum(position) space partitions of the free scalar field vacuum, in the normal-mode ground state of ion motion, $\hat{a}_\alpha|0\rangle^{\otimes N} = 0$, partitions among the local modes are naturally entangled.    
    }
    \label{fig:AxialMotionDiagram}
\end{figure}
A visualization of the ion chain and trap geometry is provided in Fig.~\ref{fig:AxialMotionDiagram}, showing the global normal modes and their linear combinations that produce the basis of local axial motional modes.

In the global normal mode vacuum, $\hat{a}_\alpha|0\rangle = 0$, with two-point expectation values of
$\bar{\omega}_\alpha\langle 0| \hat{\bar{\xi}}_\alpha \hat{\bar{\xi}}_\beta |0\rangle
= \frac{1}{\bar{\omega}_\alpha} \langle 0| \hat{\bar{\nu}}_\alpha \hat{\bar{\nu}}_\beta|0\rangle = \frac{\delta_{\alpha\beta}}{2} $,
the two-point functions of the local oscillators are
\begin{align}
\frac{\langle 0 | \hat{\phi}_i \hat{\phi}_j |0\rangle}{\ell_{nm}^2} &= \frac{1}{2} \sum_{\alpha=1}^N \frac{\left(\mathbf{e}^{-1}\right)_{i\alpha} \left( \mathbf{e}^{-1} \right)_{j\alpha} }{\bar{\omega}_\alpha}
\label{eq:localposition2point}
\\
\frac{\ell_{nm}^2}{\hbar^2}\langle 0| \hat{\pi}_i \hat{\pi}_j |0\rangle &= \frac{1}{2} \sum_{\alpha=1}^N \left(\mathbf{e}^{-1}\right)_{i\alpha} \left( \mathbf{e}^{-1} \right)_{j\alpha} \bar{\omega}_\alpha
\label{eq:localmomentum2point}
 \ \ \ .
\end{align}
In the local phase space basis of $\hat{\bar{\mathbf{r}}} = \left\{ \hat{\bar{\phi}}_1, \hat{\bar{\pi}}_1, \hat{\bar{\phi}}_2, \hat{\bar{\pi}}_2, \cdots, \hat{\bar{\phi}}_N, \hat{\bar{\pi}}_N \right\}$, where the commutation relations are encapsulated in the symplectic matrix 
    $\Omega = -i \left[\hat{\bar{\mathbf{r}}}, \hat{\bar{\mathbf{r}}}^T \right] = \bigoplus_{j = 1}^N i Y$ with $Y$ being the second Pauli matrix, the dimensionless ground state covariance matrix (CM) is
\begin{equation}
\bar{\sigma}_{i,j} = \langle 0| \left\{ \hat{\bar{r}}_{i}- \langle\hat{\bar{r}}_i\rangle, \hat{\bar{r}}_{j}- \langle\hat{\bar{r}}_j \rangle\right\}_+ |0\rangle \ \ \ .
\label{eq:cmformal}
\end{equation}
Examples of these covariance matrices for the local axial motional modes in small ion chains are provided in Appendix~\ref{app:smallexamples}.
Because the vacuum expectation values of position-momentum anticommutators vanish,
\begin{equation}
\langle 0 | \left\{ \hat{\phi}_i, \hat{\pi}_j \right\}_+ |0\rangle = 0 ,
\end{equation}
the two point functions in Eqs.~\eqref{eq:localposition2point} and~\eqref{eq:localmomentum2point} fully determine the CM, and thus the entanglement properties of the local motional modes in the leading-order Gaussian ground state.

The independence of the dimensionless CM, Eqs.~\eqref{eq:localposition2point}-\eqref{eq:cmformal}, with respect to both trap and ion experimental parameters indicates a universality of the harmonic approximation (Eq.~\eqref{eq:harmonicapprox}) of the trapped ion local motional mode correlation structure.
With a single dimensionless CM to describe all ion choices and quadratic trap strengths, this universality results in leading-order insensitivity to experimental implementations.
This may prove useful for error robustness, or, conversely, yield challenges as all modifications to the CM, e.g., for optimizing connections to the scalar field vacuum as discussed in Sec.~\ref{sec:fidelity}, must be implemented through explicit quantum operations.
Furthermore in this context, the dimensionful CM incorporates all ion and trap parameters through the length scale $\ell_{nm}^2~=~\frac{\hbar}{\sqrt{2 m q \kappa_2}}$, which functions numerically as a global set of single-mode squeezing parameters with zero entanglement power, i.e.,  scaling the $\langle \hat{\boldsymbol{\phi}}\hat{\boldsymbol{\phi}}\rangle$ expectation values inversely from those of conjugate-momentum operators $\langle \hat{\boldsymbol{\pi}}\hat{\boldsymbol{\pi}}\rangle$, c.f., Eqs.~\eqref{eq:localposition2point} and~\eqref{eq:localmomentum2point}.
As such, leading-order entanglement quantities among local modes with a quadratic trapping potential are pure numbers independent of the variables of experimental design.

\section{Entanglement Resource}
\label{sec:entanglementresource}

To assess the entanglement naturally distributed among the local motional modes, consider entanglement properties between two disjoint regions, $A$ and $B$, each composed of $d$ contiguous ions  and separated by $\tilde{r}$ ions, as shown at the center of Fig.~\ref{fig:AxialMotionDiagram}, for example.
For this bipartite state, three logarithmic negativity values  are of current interest: that when the local modes outside the regions are traced, resulting in an $A$\nobreakdash-$B$ mixed state, and those when the local modes outside the regions are projectively  measured in the $\phi$- or $\pi$\nobreakdash-basis~\footnote{See Ref.~\cite{Klco:2023ojt} for further technical details}, resulting in $A$\nobreakdash-$B$ pure states.
With differing levels of classical information provided from possible external observers,  these measures quantify here the operational amount of entanglement that could be consolidated into two-mode pairs spanning the disjoint regions~\cite{Williamson1936,2003PhRvA..67e2311B,Klco:2021cxq}.

\begin{figure}
\includegraphics[width=0.45\textwidth]{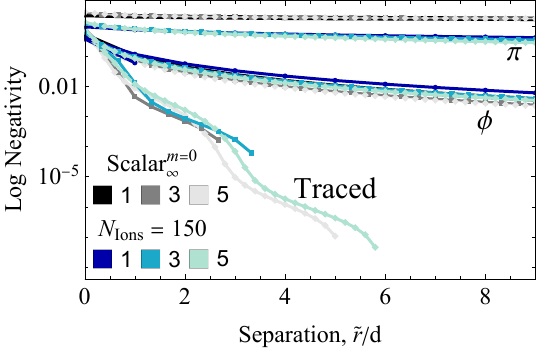}
\caption{Logarithmic negativity between two regions (of sizes $d = 1, 3, 5$ modes each) as a function of their separation in one spatial dimension.
In the ion application (solid), each mode is a local axial motional mode, with the pair of regions centered within a 150-ion chain.
In the scalar field application (dashed), each mode is a field lattice site within an infinite volume in the massless regime ($m = 10^{-10}$).
Each system is shown when the volume external to the regions is traced or measured in the local position ($\phi$) or local conjugate-momentum ($\pi$) basis.
}
\label{fig:logneg}
\end{figure}

As demonstrated in Fig.~\ref{fig:logneg} for disjoint regions of local motional modes separated across the center of a 150-ion chain, the logarithmic negativity  decays exponentially with the  separation between regions
upon tracing of the modes in the remaining chain~\cite{Marcovitch:2008sxc,Klco:2020rga,Klco:2021biu}.
 Originating from a system with  field(conjugate-momentum) two-point functions decaying logarithmically(polynomially) with spatial separation, the emergence of this strong decay is a result of the classical noise introduced in the tracing procedure~\cite{Klco:2023ojt}, i.e., resulting from the loss of information about an entangled quantum system.
These features are common to the massless scalar field vacuum, for which Fig.~\ref{fig:logneg} shows the analogous calculation in grayscale for reference.
Beyond the magnitude and decay, the calculation in Fig.~\ref{fig:logneg} also indicates close agreement in the separation at which the classical noise causes these regions of finite pixelation to transition into a regime of vanishing negativity~\cite{Audenaert:2002xfl,2004PhRvA..70e2329B,kofler2006entanglement,Marcovitch:2008sxc,Calabrese:2009ez,Calabrese:2012ew,Calabrese:2012nk,MohammadiMozaffar:2017nri,Coser_2017,Javanmard:2018xte,Klco:2019yrb,DiGiulio:2019cxv,Klco:2020rga,Klco:2021biu} that is also separable~\cite{2001PhRvL..87p7904G,Klco:2021biu}.
With no free parameters,  the externally traced logarithmic negativity is remarkably similar between these two distinct physical systems, one at the center of a quadratically trapped ion chain and the other within an infinite volume of a free massless scalar field vacuum.

Upon measurement of the external modes in the local position $\phi$-basis, Fig.~\ref{fig:logneg} shows that the two systems again exhibit comparable underlying entanglement resources.
However, upon measurement of the external modes in the conjugate-momentum $\pi$-basis~\footnote{Explicit forms of the classical noise removed upon measurement in each basis are provided in Ref.~\cite{Klco:2023ojt}.}, the distinction between these two physical systems becomes apparent.
This difference can be understood as a result of the field gradients in the scalar field and the quadratically balanced Coulomb repulsion in the trapped ion chain distributing inequivalent vacuum entanglement.

Beyond the magnitudes of these entanglement measures, the local motional modes are found to have further structural entanglement features in common with the free lattice scalar field.
Though the spatial parity symmetry, present when regions are centered within the ion chain, is generally insufficient for the techniques of Ref.~\cite{Klco:2021cxq} to ensure consolidation of negativity into a separable set of two-mode entangled pairs governed by the partially transposed (PT) symplectic eigenvalues,  it is found that the local motional modes within the trapped ion chain do share this consolidation feature with the free scalar field vacuum.
As such, the PT symplectic eigenvectors provide guidance for locally transforming the accessible many-body entanglement between disjoint regions of ions into a series of two-mode squeezed states.
The availability of this two-mode organization of many-body mixed-state entanglement may prove to be useful experimentally in transferring entanglement from the local motional modes into discrete ion internal energy levels, as first proposed in Ref.~\cite{Retzker2005:gor} and further discussed in Section~\ref{sec:transferResource}.

\section{Preliminary Applications}
\label{sec:prelimapps}
\subsection{Vacuum State Preparation}
\label{sec:fidelity}

For digital quantum algorithms, studies establishing the efficiency (in e.g., particle number, coupling strength, etc.) of simulating scattering in interacting quantum fields~\cite{Jordan:2011ci,Jordan:2011ne} show that the dominant use of quantum computational resources is expected to be required not in the real-time dynamics, but in the process of initial state preparation, e.g., of the quantum field vacuum.
From a complementary perspective,  exploring the low energy subspace and preparing states within it from a trivial tensor-product fiducial state can be beyond the expected asymptotic efficiency of quantum hardware and software, even for few-body spatially local Hamiltonians~\cite{kempe_kitaev_regev_2004,oliveira2008complexity,freedman2014quantum,Anshu:2022hsn}.
These considerations, as well as the presence of phase transitions in lattice gauge theories, motivate the exploration of diverse methods of entangled quantum field vacuum state preparation~\cite{kitaev2009wavefunction,Kaplan:2017ccd,HamedMoosavian:2017koz,Lee:2019zze,Moosavian:2019rxg,Klco:2019yrb,Choi:2020pdg,Buser:2020uzs,Klco:2020aud,LinTong2020,Deliyannis:2021che,Ciavarella:2021lel,Funcke:2021aps,Bagherimehrab:2021xlp,Stetcu:2022nhy,Farrell:2023fgd,Cohen:2023dll}, with strategies ranging from variational techniques and guidance by tensor networks to adiabatic flows and the design of reservoirs for dynamical cooling procedures.
To this vacuum state preparation literature, the present paper introduces the possibility of replacing the trivial-complexity fiducial state with an experimentally available state of naturally distributed entanglement.

When generating entanglement through interleaved layers of few-body operators, entanglement is created with an expansion rate governed by the Lieb-Robinson bound~\cite{Lieb1972,2006JSP...124....1N}.
However, when many-body entanglement resources are naturally present in a quantum architecture, e.g., the local motional modes of a trapped ion chain, actively incorporating them in algorithmic design can yield alternatives to propagation-based entanglement distribution.
Analogous to the way in which providing a computational framework with quantum degrees of freedom allows more efficient representation and imitation of quantum systems themselves~\cite{Feynman1982}, providing a computational framework with access to naturally distributed arrays of entangled degrees of freedom could provide similar advantages for quantum field simulations.

For scalar field applications suggested by the entanglement observations above, consider quantifying the provided leading-order approximation with the quantum fidelity~\cite{UHLMANN1976273,Jozsa1994,Fuchs:1995mk,Nielsen2010-ga},
$\mathcal{F}(\rho_1, \rho_2) = \mathcal{F}(\rho_2, \rho_1) = \text{tr} \left( \sqrt{ \sqrt{\rho_1} \rho_2 \sqrt{\rho_1}}\right) $,
where $\rho_{1,2}$ are the density matrices of local axial motional modes in the center of an ion chain and of CV lattice sites within an infinite-volume scalar field.
The fidelity saturates to unity when the two density matrices are equivalent,  vanishes when they are composed of orthogonal subspaces, and may be directly related to metrics providing a formal sense of distance between quantum states~\cite{HELSTROM1967101,Bures1969}.
Furthermore, many preparation algorithms rely upon the availability of an input initial state with non-vanishing fidelity with respect to a desired state, impacting both figures of merit and necessary statistics.

\begin{figure}
\includegraphics[width=0.45\textwidth]{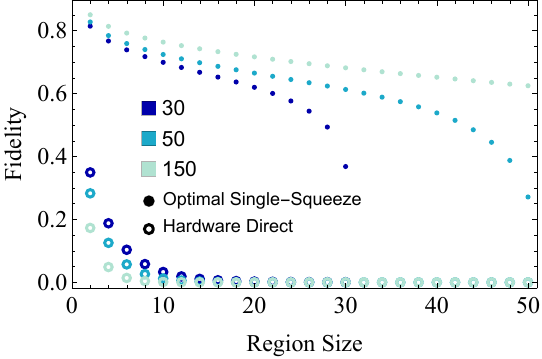}
\caption{Fidelity between the central local axial modes of trapped ion chains cooled to their motional ground state and a same-size region of the infinite-volume, lattice scalar field vacuum in the massless regime ($m = 10^{-10}$).
Shown for regions of up to 50 local modes/lattice sites for 30-, 50-, and 150-ion chains, a one-parameter optimization over a globally implemented single-mode squeezing parameter allows the creation of approximations to the latticized field vacuum with enduring fidelity.
 }
\label{fig:fidelities}
\end{figure}

For mixed Gaussian CV states, Ref.~\cite{Banchi2015} provides several techniques for calculating the fidelity from the set of first and second moments of each Gaussian state~\footnote{For example,  when the first moments of both states are vanishing, the formulation~\cite{Banchi2015},
\begin{equation}
\mathcal{F}(\sigma_1, \sigma_2) = \left(\frac{\det \left( 2 \left[\sqrt{ \mathbb{I} + \frac{\left(V\Omega\right)^{-2}}{4}}+ \mathbb{I}\right]V \right)}{\det \left( \sigma_1 + \sigma_2\right)}   \right)^{\frac{1}{4}}\ \ \ ,
\end{equation}
with
\begin{equation}
V = \Omega^T(\sigma_1+\sigma_2)^{-1}\left( \frac{\Omega}{4} + \sigma_2 \Omega \sigma_1\right) \ \ \ ,
\end{equation}
is straightforward to compute from the covariance matrices in the present harmonic approximation of the combined ion trapping and Coulomb potentials.}.
A fidelity analysis is shown in Fig.~\ref{fig:fidelities} between the local modes of the trapped-ion motional ground state and the free, massless, lattice scalar field vacuum.
Unfortunately, approximate quantum state preparations commonly yield fidelities that decay exponentially with increasing system size, becoming rapidly impractical for large-scale applications.
As shown by the open points in Fig.~\ref{fig:fidelities}, such catastrophic loss of fidelity occurs for the ion local motional modes naturally produced in the hardware.
However, inspired by the similar distribution of disjoint two-region entanglement in both systems demonstrated in Fig.~\ref{fig:logneg},  performing a one-parameter optimization of a single-mode squeezing applied to each local motional mode (a unitary with zero entanglement power) results in substantial fidelities even for large systems.
The values of this squeezing parameter as a function of region and chain size are recorded in the numerical tables in Appendix~\ref{app:numericalTables}.
For chain lengths comparable to the field region size, these squeezing magnitudes are realistic given current capabilities and ongoing developments in the experimental squeezing of ion motional modes~\cite{Heinzen1990,Cirac1993PhysRevLett.70.556,Meekhof1996,Alonso2013,Kienzler2015Sci,Kienzler2016,Burd2019,Sutherland:2021ifb}.
Note that the universality of the dimensionless CM, Eqs.~\eqref{eq:localposition2point}-\eqref{eq:cmformal}, indicates that this squeezing must be introduced through external quantum operations on each local motional mode.

It is encouraging that a simply calculated set of local squeezings can produce a state of local motional modes with similar disjoint entanglement structure and  $\mathcal{O}(1)$ fidelity with respect to large instances of the massless lattice scalar field vacuum.
The remaining fundamental distinctions between the small oscillation approximation of ion local motional modes and the scalar vacuum suggest that the calculated fidelity will not asymptote to unity for larger ion chains.
However, note that this calculation considered only the flexibility of one single-mode squeezing parameter;  for a given system, all entanglement features are thus generated only by the natural ion local motional modes themselves.
As such, this calculation is best interpreted as a leading-order result, subject to systematic improvement upon the action of a broader array of CV quantum circuitry.

These preliminary results motivate dedicated future inquiry exploring the replacement of canonical tensor-product initial states by experimentally convenient states of naturally distributed entanglement.

\subsection{Transferring Distributed Entanglement}
\label{sec:transferResource}

For purposes of simulation with finite-dimensional quantum systems or for broader computational applications of spatially distributed quantum correlations, it is of interest to consider transferring the entanglement among local CV motional modes into the Hilbert space composed of the associated ions' internal energy levels.
This process was first proposed as an opportunity for experimentally detecting vacuum entanglement and for entangling trapped ions with quantum gates utilizing only local phase-space excitations~\cite{Retzker2005:gor}.
The local motional modes then provide an entangled fabric of spatially distributed entanglement that may be strategically accessed.
That the naturally available entanglement structure is remarkably similar to that of a simple latticized quantum field suggests further possible applications from quantum simulation of Standard Model physics to large-scale quantum error correction.
Motivated by recent theoretical and experimental advances in qudit-based quantum information processing~\cite{Gokhale:2019asy,JiangLow:2019prac,Blok:2020may,Ciavarella:2021nmj,Kurkcuoglu:2021dnw,Chi:2022pro,Gustafson:2022xlj,Goss2022,Gonzalez-Cuadra:2022hxt,Zache:2023cfj,Popov:2023xft,Low:2023dlg,Meth:2023wzd}, the following section details one important consideration affecting the magnitude and efficiency of CV-to-qudit entanglement transfer in the context of trapped ions.

There are many strategies for transferring entanglement from CV local motional modes to finite dimensional qudit systems~\cite{Son2002entTransfer,Kraus2004entDist,Paternostro2004,Retzker2005:gor,Serafini2006entTransfer,Casagrande2007}.
One basic paradigm involves writing the CV system in an indexed orthonormal basis and performing an approximate swapping operation between $n$ basis states and the $n$-dimensional qudit system.
For example, Ref.~\cite{Retzker2005:gor} presents an approximation of such a SWAP operator for an expression of the CV local motional modes in the Fock basis, transferring the quantum information present in the low-energy Fock space to qubits of ion internal energy levels.
Generalizing to qudits, the population of the CV system in states external to the chosen subspace (e.g., above the lowest energy Fock states)  yields excess population in the qudit initial state, reducing the transferred entanglement.
For the Fock-space transfer, this external-population figure of merit is shown in Fig~\ref{fig:subspaceprobs2ions} of Appendix~\ref{app:fockspace} for the local motional modes of a two-ion chain as a function of increasing qudit dimension.

Although the entanglement between two Hilbert spaces is not impacted by local operations, the entanglement accessible from within a Hilbert space subset can be highly sensitive to local restructuring.
As such, local CV operations applied prior to qudit extraction into the ion internal states can impact the probability in the qudit subspace and thus the transferred entanglement.
Figure~\ref{fig:subspaceprobs2ions} indicates that first including a pair of single-mode squeezings to transform the local motional mode CM into normal form with balanced $\langle \boldsymbol{\phi}\boldsymbol{\phi}\rangle$ and $\langle \boldsymbol{\pi}\boldsymbol{\pi}\rangle$ matrix elements reduces the population external to the low-energy Fock states by several orders of magnitude, with the impact increasing with increasing qudit dimension.
By calculating directly the motional mode CM naturally produced in the ion chain, such operations can be designed to optimize the localization of the ion modes in an experimentally convenient orthonormal basis for entanglement transfer.

In designing local operators for more than two ions to optimize this form of CV to finite-dimensional entanglement transfer, the observation discussed at the end of Section~\ref{sec:entanglementresource}---that the ion local motional modes in this context satisfy requirements allowing the techniques of Ref.~\cite{Klco:2021cxq} to reorganize the available entanglement between regions---provides an additional advantage.
Such consolidation, informed by the eigensystem of the partially transposed $2d$-ion  mixed-state, reorganizes the bipartite entanglement between disjoint many-body field or ion-chain regions into a tensor-product series of $(1_A \times 1_B)$ entangled pairs.
As such, this reorganization pre-conditions the CV entanglement to be conducive to subsequent entanglement transfer, ultimately allowing this entanglement to be consolidated into fewer ion pairs and improving the entanglement yield.

\section{Summary and Outlook}
Considering entanglement structure from multiple operational perspectives can serve to quantify resources available for quantum information protocols and to guide the design of quantum simulations.
This paper considers whether the local axial motional modes of a simple trapped ion system can serve as a useful starting point for generation of entangled vacuum states of the massless scalar field.
Recognizing that several region-region entanglement measures exhibit strong similarities between the two systems, analyses of fidelities and localization in Fock-space constitute initial steps toward utilizing naturally distributed entanglement resources for computational applications.
This perspective highlights common elements among quantum simulation, sensing, and networking as quantum fields are considered for incorporation directly into quantum architectures.

The present analysis, however, only modestly addresses a subset of the important questions driving research toward quantum technology goals.
For scientific application to quantum state preparation, the approximate correspondence observed in Section~\ref{sec:fidelity} must be promoted to a systematically improvable protocol, possibly through the introduction of a hierarchy of higher-complexity quantum operations~\cite{Klco:2019yrb,Klco:2020aud}.
Furthermore, future development toward quantum simulation applications should be made beyond the idealized motional vacuum to include experimentally attainable quantum states after cooling procedures minimize, but do not eliminate, collective phonon numbers.

For the application of Section~\ref{sec:transferResource}, further work is needed to determine practical techniques for implementing an approximate swapping operator between the local motional modes and ion internal levels, including the assessment of many choices of basis and the design of pre-conditioning procedures to strategically organize entanglement.
Better understanding of successful procedures for imaging naturally distributed CV entanglement to finite dimensions may further inform the design of entanglement-efficient bosonic truncation and digitization schemes, e.g., in designing computational bases for the practical quantum simulation of quantum field theories~\cite{PhysRevA.73.022328,Zohar:2013zla,Martinez:2016yna,Raychowdhury:2018osk,Klco:2018zqz,Alexandru:2019nsa,Klco:2019evd,Kreshchuk:2020dla,Davoudi:2020yln,Haase:2020kaj,Ciavarella:2021nmj,Bauer:2021gek,Wiese:2021djl,Aidelsburger:2021mia,Gonzalez-Cuadra:2022hxt,Pardo:2022hrp,Kadam:2022ipf,Bauer2023,DAndrea:2023qnr,Funcke:2023jbq}.

The present connection between the scalar field vacuum and local motional modes of a trapped-ion chain, guided by entanglement structure, brings present research still closer to the vision of quantum simulation~\cite{Feynman1982}---leveraging the mutual intersimulatability of diverse quantum systems, even those widely separated in natural energy scales, e.g., the connection between atomic and subatomic physics arising in present quantum simulations of quantum chromodynamics.

\vspace{0.2cm}
\begin{acknowledgments}
We thank R.~Blatt, C.~Noel, G.~Pagano, M.~J. Savage, and P.~Zoller for inspiring discussions.
The creative initiation of several aspects of this work occurred during the workshop \emph{At the Interface of Quantum Sensors and Quantum Simulation} at the InQubator for Quantum Simulation (IQuS) hosted by the Institute for Nuclear Theory (INT).
We thank R.~Blatt and P.~Zoller for challenging the organizers with homework throughout this workshop.
D.H.B. is supported in part by NSF Nuclear Physics grant PHY-2111046.
\end{acknowledgments}

\bibliography{biblio}

\onecolumngrid
\appendix

\section{Small-Chain Examples}
\label{app:smallexamples}

With a quadratic trapping potential, the equilibrium positions for two- and three-ion chains can be evaluated exactly  to be
\begin{equation}
\mathbf{z}^0_{N = 2} = \left\{ - \frac{2^{1/3} \ell_{\mu m}}{2}, \frac{2^{1/3} \ell_{\mu m}}{2} \right\}
\quad , \quad
\mathbf{z}_{N=3}^0 = \left\{ - \frac{10^{1/3} \ell_{\mu m}}{2}, 0, \frac{10^{1/3} \ell_{\mu m}}{2}\right\} \ \ \ ,
\end{equation}
where $\ell_{\mu m} = \left( \frac{q }{8 \pi \epsilon_0 \kappa_2} \right)^{1/3}$ is a characteristic ion-separation length scale.
The resulting frequencies and normal modes are
\begin{equation}
\bar{\boldsymbol{\omega}} = \frac{\boldsymbol{\omega}}{\omega_z} =  \left\{ 1, \sqrt{3} \right\}
\quad , \quad
 e = \begin{pmatrix}
\mathbf{e}_{1} \\
\mathbf{e}_{2}
\end{pmatrix} =  \frac{1}{\sqrt{2}} \begin{pmatrix}
1 & 1 \\
-1 & 1
\end{pmatrix} \ \ \
\end{equation}
for a two-ion chain and
\begin{equation}
\bar{\boldsymbol{\omega}} = \frac{\boldsymbol{\omega}}{\omega_z} =  \left\{ 1, \sqrt{3}, \sqrt{\frac{29}{5}} \right\}
\quad , \quad
 e = \begin{pmatrix}
\mathbf{e}_1 \\
\mathbf{e}_2 \\
\mathbf{e}_3
\end{pmatrix} = \begin{pmatrix}
\frac{1}{\sqrt{3}} & \frac{1}{\sqrt{3}} & \frac{1}{\sqrt{3}} \\
-\frac{1}{\sqrt{2}} & 0 & \frac{1}{\sqrt{2}} \\
\frac{1}{\sqrt{6}} & - \frac{2}{\sqrt{6}} & \frac{1}{\sqrt{6}}
\end{pmatrix} \ \ \
\end{equation}
for a three-ion chain.
Interestingly, the ground state center-of-mass motion frequency and gap to the excited state breathing mode are known to be independent of the number of ions in the chain~\cite{Cirac:1995zz}.

Utilizing Eqs.~\eqref{eq:localposition2point}-\eqref{eq:cmformal}, the dimensionless covariance matrices for two and three ions are directly calculated to be
\begin{equation}
\bar{\sigma}_{N = 2} = \frac{1}{2}
\left(
\begin{array}{cccc}
 \frac{3+\sqrt{3}}{3}  & 0 & \frac{3-\sqrt{3}}{3}  & 0 \\
 0 & 1+\sqrt{3} & 0 & 1-\sqrt{3} \\
 \frac{3-\sqrt{3}}{3} & 0 & \frac{3+\sqrt{3}}{3} & 0 \\
 0 & 1-\sqrt{3} & 0 & 1+\sqrt{3} \\
\end{array}
\right) \quad , \quad \sigma_{N = 2} = \frac{\ell_{nm}^2}{2}
\left(
\begin{array}{cccc}
 \frac{3+\sqrt{3}}{3}  & 0 & \frac{3-\sqrt{3}}{3}  & 0 \\
 0 & \frac{\hbar^2\left(1+\sqrt{3}\right)}{\ell_{nm}^4}  & 0 & \frac{\hbar^2 \left(1-\sqrt{3} \right)}{\ell_{nm}^4} \\
 \frac{3-\sqrt{3}}{3} & 0 & \frac{3+\sqrt{3}}{3} & 0 \\
 0 & \frac{\hbar^2 \left(1-\sqrt{3}\right)}{\ell_{nm}^4} & 0 & \frac{\hbar^2 \left(1+\sqrt{3} \right)}{\ell_{nm}^4} \\
\end{array}
\right)
\label{eq:cmN2}
\end{equation}
and
\begin{equation}
\bar{\sigma}_{N = 3} = \scalemath{0.8}{
\left(
\begin{array}{cccccc}
 \frac{29 \sqrt{3}+\sqrt{145}+58}{174}  & 0 & \frac{1}{3}-\frac{\sqrt{\frac{5}{29}}}{3} & 0 & \frac{-29 \sqrt{3}+\sqrt{145}+58}{174}  & 0 \\
 0 & \frac{15 \sqrt{3}+\sqrt{145}+10}{30}  & 0 & \frac{5-\sqrt{145}}{15}  & 0 & \frac{-15 \sqrt{3}+\sqrt{145}+10}{30}  \\
 \frac{1}{3}-\frac{\sqrt{\frac{5}{29}}}{3} & 0 & \frac{1}{3}+\frac{2 \sqrt{\frac{5}{29}}}{3} & 0 & \frac{1}{3}-\frac{\sqrt{\frac{5}{29}}}{3} & 0 \\
 0 & \frac{5-\sqrt{145}}{15}  & 0 & \frac{2 \sqrt{145}+5}{15}  & 0 & \frac{5-\sqrt{145}}{15}  \\
 \frac{-29 \sqrt{3}+\sqrt{145}+58}{174} & 0 & \frac{1}{3}-\frac{\sqrt{\frac{5}{29}}}{3} & 0 & \frac{29 \sqrt{3}+\sqrt{145}+58}{174}  & 0 \\
 0 & \frac{-15 \sqrt{3}+\sqrt{145}+10}{30}  & 0 & \frac{5-\sqrt{145}}{15} & 0 & \frac{15 \sqrt{3}+\sqrt{145}+10}{30} \\
\end{array}
\right) }\ \ \ .
\label{eq:cmN3}
\end{equation}
As demonstrated in Eq.~\eqref{eq:cmN2}, restoring the dimensions of the CM is achieved, according to Eq.~\eqref{eq:localposition2point} and~\eqref{eq:localmomentum2point}, through insertion of $\ell_{nm}^2$ or $\frac{\hbar^2}{\ell_{nm}^2}$ on the $\langle \hat{\boldsymbol{\phi}} \hat{\boldsymbol{\phi}}\rangle$ or $\langle \hat{\boldsymbol{\pi}} \hat{\boldsymbol{\pi}}\rangle$ submatrix, respectively.
Note that the dimensionless covariance matrix of this harmonic approximation is universal, i.e., no trap or ion experimental parameters are present.
Therefore, all entanglement properties are pure numbers, independent of the specifics of the chosen ion and strength of the quadratic trapping potential.
In particular, the two-mode logarithmic negativities are found to be
\begin{align}
\mathcal{N}^{N=2}_{1|2} &= \frac{\ln 3}{\ln 16}  = 0.396241...
\\
\mathcal{N}^{N=3}_{1|2} &= \frac{\ln \left(-\frac{5220}{-145 \left(4 \sqrt{3}+9\right)-2 \sqrt{145} \left(11 \sqrt{3}+153\right)+\sqrt{435 \left(-5952 \sqrt{3}+4 \left(105 \sqrt{3}+299\right) \sqrt{145}+13571\right)}}\right)}{\ln 4} = 0.384585...
\\
\mathcal{N}^{N=3}_{1|3} &= \frac{\ln \left(\frac{5}{3} \left(49-4 \sqrt{145}\right)\right)}{\ln 16}  = 0.118608...
\end{align}
where the third mode in the $N=3$ examples has been traced.
For the two-ion chain, the single-mode entanglement entropies are found to be
\begin{multline}
E_1^{N=2} = \frac{1}{\ln 4096}  \left[\left(\sqrt{6 \left(2 \sqrt{3}+3\right)}+6\right) \ln \left(\sqrt{6 \left(2 \sqrt{3}+3\right)}+6\right) \right. \\ \left. -12 \ln 12-\left(\sqrt{6 \left(2 \sqrt{3}+3\right)}-6\right) \ln \left(\sqrt{6 \left(2 \sqrt{3}+3\right)}-6\right) \right] = 0.13618...
\end{multline}
consistent with Ref.~\cite{Retzker2005:gor}
where the two-mode state of the local motional modes was taken to be a two mode squeezed vacuum state in normal form, i.e., consistent with Eq.~\eqref{eq:cmN2} up to local operations.

\section{Local Motional Modes in Fock-space}
\label{app:fockspace}

A significant convenience of Gaussian CV quantum states is that they are completely described by a covariance matrix of second moments, $\sigma$, and a vector of first moments, $\bar{\mathbf{r}}$, in phase space.
The dimensionality of these two objects scales only linearly with the number of modes in the system, allowing states of significant size to be computationally practical.
However, transferring entanglement from CV to finite-dimensional quantum systems---e.g., from local motional degrees of freedom to internal energy levels of trapped ions as first proposed by Ref.~\cite{Retzker2005:gor}---inspires departing from the succinct phase-space language by expanding the CV state in the infinite tower of Fock states.
From this basis, Ref.~\cite{Retzker2005:gor} discussed the experimental viability of dynamically swapping the lowest-excitation motional Fock states with an internal state ion qubit.

Before evaluating the Fock basis decomposition for an arbitrary covariance matrix,  consider first the Fock representation of a two mode squeezed vacuum state~\cite{Serafini2017},
\begin{equation}
|\psi_r\rangle = e^{r \left( \hat{a}_1^\dagger \hat{a}_2^\dagger - \hat{a}_1 \hat{a}_2 \right) } |0,0\rangle  \ \ \ ,
\label{eq:tmsvsFockOp}
\end{equation}
where $\hat{a}_j$ and $\hat{a}_j^\dagger$ are bosonic creation and annihilation operators, $\left[ \hat{a}_i, \hat{a}_j^\dagger \right] = \delta_{ij}$.
As usual, the associated covariance matrix can be calculated via application of the two-mode squeezing symplectic operation on the vacuum (identity) covariance matrix,
\begin{equation}
\sigma_r = S_rS_r^T \qquad , \qquad S_r = \begin{pmatrix}
\cosh r & 0 & \sinh r & 0 \\
0 & \cosh r & 0 & -\sinh r \\
\sinh r & 0 & \cosh r & 0 \\
0 & -\sinh r & 0 & \cosh r
\end{pmatrix} \ \ \ .
\end{equation}
Evaluation of Eq.~\eqref{eq:tmsvsFockOp} can be performed by utilizing the su(1,1) algebra, with generators $K_{\pm}$ and $K_0$ satisfying the following commutation relations,
\begin{equation}
\left[ K_+, K_-\right] = -2K_0 \qquad , \qquad  \left[ K_0, K_\pm \right] = \pm K_{\pm} \ \ \ .
\end{equation}
Transformations in this context can be simplified by using so-called disentanglement theorems~\cite{Gilmore1974,Santiago_1976},
\begin{equation}
e^{v_+ K_+ + v_- K_- + v_0 K_t} = e^{t_+ K_+} e^{\left(\ln t_0\right) K_0} e^{t_- K_-} \ \ \ .
\end{equation}
Determination of the coefficients $\vec{t}$ from those of $\vec{v}$ can be achieved by utilizing a small-dimensional representation of the generators---e.g., $K_+ = \begin{pmatrix}
0 & 1 \\
0 & 0
\end{pmatrix}$,  $K_- = \begin{pmatrix}
0 & 0 \\
-1 & 0
\end{pmatrix}$, and $K_0 = \frac{1}{2}Z$, with $Z$ being the third Pauli matrix---and equating each matrix element of the two transformations.
The resulting coefficients are
\begin{align}
t_0 &= \left( \cosh(f) - \frac{v_0 }{2 f} \sinh(f)\right)^{-2} \ \ \ , \\
t_{\pm} &= \frac{v_\pm \sinh(f)}{f \Big\lvert\cosh(f) - \frac{v_0 }{2 f} \sinh(f) \Big\rvert } \ \ \ ,
\end{align}
with parameter $f^2 = \frac{1}{4} v_0^2 - v_+ v_-$.  
For application to the two-mode squeezed vacuum state, one may identify $K_+=\hat{a}_1^\dagger \hat{a}_2^\dagger$ and  $K_- = \hat{a}_1 \hat{a}_2$, leading to  $K_0 = \frac{1}{2} \left( \hat{a}_1^\dagger \hat{a}_1 + \hat{a}_2^\dagger \hat{a}_2 + 1\right)$.
The squeezing operation may thus be written in the well-known Fock basis form,
\begin{align}
e^{r \left( \hat{a}_1^\dagger \hat{a}_2^\dagger - \hat{a}_1 \hat{a}_2 \right) } |0,0\rangle &= e^{\tanh(r) \hat{a}_1^\dagger \hat{a}_2^\dagger} e^{-\left(\ln \cosh{r}\right) \left( \hat{a}_1^\dagger \hat{a}_1 + \hat{a}_2^\dagger \hat{a}_2 + 1\right)} e^{-\tanh(r) \hat{a}_1 \hat{a}_2} |0,0\rangle  \label{eq:disentexp}\\
& = \frac{1}{\cosh(r)} \sum_{n = 0}^\infty \tanh^n(r) |n, n\rangle \ \ \  \\
&\equiv  \sqrt{1-e^{-2\beta}} \sum_{n = 0}^{\infty} e^{-\beta n}|n,n\rangle \ \ \ ,
\label{eq:tmsvsFockbasis}
\end{align}
with $\beta \equiv - \ln \tanh(r)$ commonly introduced.
When acting on the vacuum state in Eq.~\eqref{eq:disentexp}, only the identity survives from the first exponential on the right, the second exponential becomes simply a normalization factor, and the last exponential produces a tower of states with equal occupation and dominant support in the space of low excitations.
As stronger squeezing is applied, the distribution in Fock space becomes increasingly delocalized and the entanglement increases.  For example, the logarithmic negativity, which is a necessary and sufficient measure of entanglement in this two-mode context, is directly proportional to the squeezing parameter as~$\mathcal{N} = \frac{2 |r|}{\ln 2}$.

For general Gaussian states, it is valuable to calculate Fock-basis density matrix elements directly.
Each Fock state, $|\mathbf{n}\rangle = \left(n_1, \cdots, n_N\right)$, is labeled by a vector of numbers corresponding to the quantized excitations present in each of the $N$ modes.
With diagonal elements of the density matrix, $\langle \mathbf{n}|\rho|\mathbf{n}\rangle$, commonly calculated for applications in Gaussian boson sampling~\cite{Hamilton2017gbs,Kruse2019dsgbs}, computational techniques for the generic matrix element, $\langle \mathbf{m}|\rho|\mathbf{n}\rangle$ applicable to mixed quantum states have been developed recently~\cite{Quesada2019sim}.
For states with $\bar{\mathbf{r}}=\mathbf{0}$, the techniques of Ref.~\cite{Quesada2019sim} are summarized by the relations,
\begin{equation}
	\langle \mathbf{m}|\rho |\mathbf{n}\rangle = \frac{\text{haf}\left(\tilde{A}\right)}{\sqrt{\det\left(\sigma_Q\right)\prod_{j=1}^N m_j! n_j!}}  \qquad , \qquad \tilde{A} = \bar{A}-\text{diag}(\bar{A}) \quad , \quad A = X\left(\mathbb{I}_{2N} - \sigma_Q^{-1}\right)
\end{equation}
with
\begin{equation}
X = \begin{pmatrix}
0 & \mathbb{I}_{N} \\
\mathbb{I}_{N} & 0
\end{pmatrix} \quad , \quad \sigma_Q = \frac{1}{2} \left(U_{\hat{a}}\sigma_{xxpp}U_{\hat{a}}^\dagger + \mathbb{I}_{2N}\right) \quad , \quad  U_{\hat{a}} = \frac{1}{\sqrt{2}} \begin{pmatrix}
\mathbb{I}_N & i \mathbb{I}_N \\
\mathbb{I}_N & -i\mathbb{I}_N
\end{pmatrix}
\end{equation}
where $U$ is the matrix transferring phase space operators into bosonic creation and annihilation operators and $\sigma_{xxpp}$ is the position-momentum reordered covariance matrix according to the basis, $\hat{\mathbf{r}}' = \left\{ \hat{\phi}_1,  \hat{\phi}_2,  \cdots, \hat{\phi}_N, \hat{\pi}_1, \hat{\pi}_2, \cdots, \hat{\pi}_N \right\}$.
Remaining to be discussed are the Hafnian and $\bar{A}$.
As usual, the Hafnian is a matrix property that can be calculated as
\begin{equation}
\text{haf}(B) = \sum_{\vec{j}\in \mathcal{M}} B_{j_1,j_2} \cdots B_{j_{n-1},j_n}
\end{equation}
where $\mathcal{M}$ is the set of all ways to partition the $n$ elements of $B$ into non-repeating pairs.  For example, $\text{haf}\left( B_{4\times 4}\right) = B_{12}B_{34} + B_{13}B_{24} + B_{14}B_{23}$, and the Hafnian of an odd-dimensioned matrix is zero.
The Hafnian of a zero-dimensional matrix is set to be unity.
While this computational form of the Hafnian makes manifest its combinatoric nature, note that Wick's theorem also allows $\text{haf}\left(B\right)$ to be calculated as a Gaussian expectation value of the product of all phase space operators subject to a covariance matrix governed by $B$, and first moment displacements, $\text{diag}(B)$.
Finally, the matrix~$\bar{A}$ is constructed from $A$ through  repetition of its $j^{\text{th}}$ rows and columns $n_j$($m_j$) times for the first(second) $N$ dimensions to create a matrix of dimension, $\left[ \bar{A} \right] = \sum_j n_j+m_j$, scaling with the total number of Fock space excitations involved in the matrix element.

Applying this formalism to the trapped ion local motional modes, consider the harmonic limit (Eq.~\eqref{eq:Hbar2}) of the two-ion system, characterized by the covariance matrix of Eq.~\eqref{eq:cmN2}.
In the Fock basis, the density matrix elements for the first three levels in the basis of $\left\{|00\rangle, |01\rangle, |02\rangle, |10\rangle, |11\rangle, |12\rangle, |20\rangle, |21\rangle, |22\rangle\right\}$ will read
\begin{equation}
\bar{\rho}_{(0,1,2)} =
\scalemath{0.95}{
\left(
\begin{array}{ccccccccc}
 0.963 & 0 & \cellcolor{TealBlue!25}-0.0913 & 0 & 0.129 & 0 & \cellcolor{TealBlue!25}-0.0913 & 0 & 0.0259 \\
 0 & 0 & 0 & 0 & 0 & 0 & 0 & 0 & 0 \\
 \cellcolor{TealBlue!25}-0.0913 & 0 & \cellcolor{TealBlue!25}0.00865 & 0 & \cellcolor{TealBlue!25}-0.0122 & 0 & \cellcolor{TealBlue!25}0.00865 & 0 & \cellcolor{TealBlue!25}-0.00246 \\
 0 & 0 & 0 & 0 & 0 & 0 & 0 & 0 & 0 \\
 0.129 & 0 & \cellcolor{TealBlue!25}-0.0122 & 0 & 0.0173 & 0 & \cellcolor{TealBlue!25}-0.0122 & 0 & 0.00348 \\
 0 & 0 & 0 & 0 & 0 & 0 & 0 & 0 & 0 \\
 \cellcolor{TealBlue!25}-0.0913 & 0 & \cellcolor{TealBlue!25}0.00865 & 0 & \cellcolor{TealBlue!25}-0.0122 & 0 & \cellcolor{TealBlue!25}0.00865 & 0 & \cellcolor{TealBlue!25}-0.00246 \\
 0 & 0 & 0 & 0 & 0 & 0 & 0 & 0 & 0 \\
 0.0259 & 0 &\cellcolor{TealBlue!25} -0.00246 & 0 & 0.00348 & 0 & \cellcolor{TealBlue!25}-0.00246 & 0 & 0.000698 \\
\end{array}
\right)}
 \ \ \ .
\end{equation}
Note that, as a subset, this portion of the density matrix is neither pure nor normalized.
The presence of the highlighted elements is a clear indication that the two-ion state of local motional modes is not simply a two-mode squeezed vacuum state, Eq.~\eqref{eq:tmsvsFockbasis}, as presented in Ref.~\cite{Retzker2005:gor}.  However, one can transform the two-ion local motional modes into a two-mode squeezed vacuum state through local symplectic operations, leading to agreement with the reported entanglement properties for the two-ion system.

\begin{figure}
\centering
\includegraphics[width=0.95\textwidth]{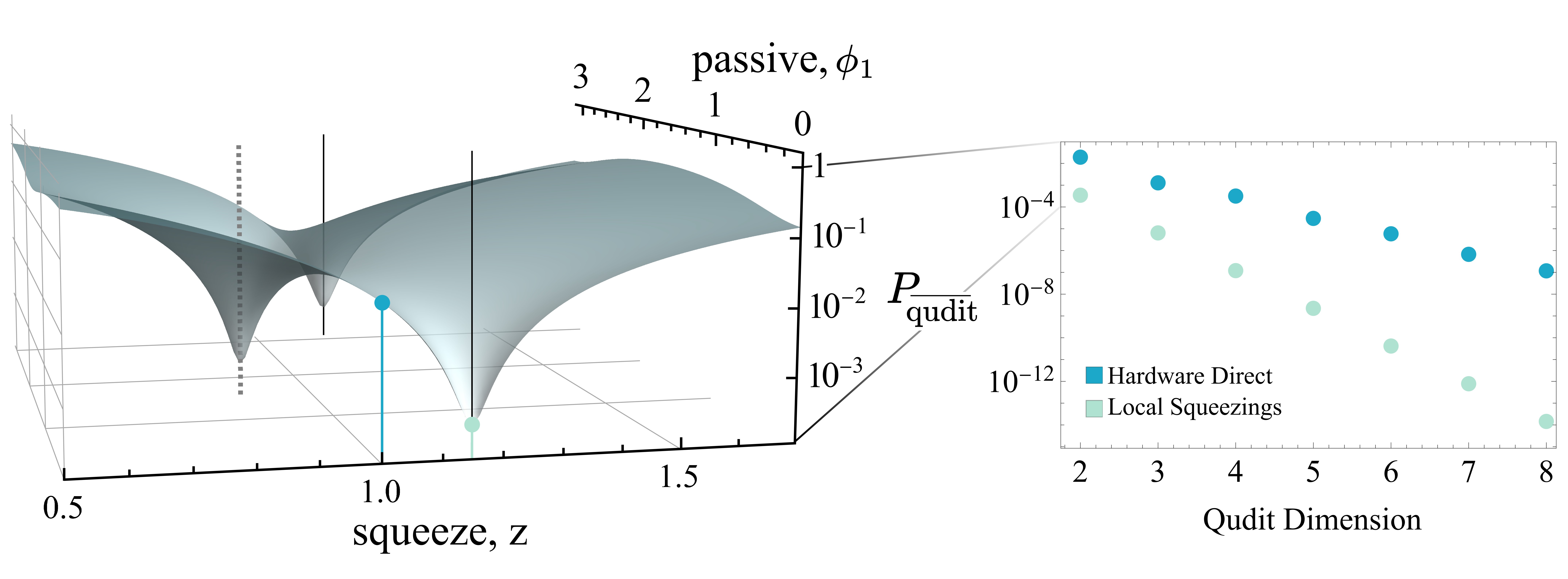}
\caption{For the local motional modes of a two-ion chain, the probability outside the lowest two-level Fock subspace ($P_{\overline{\text{qubit}}}=1-P_{\text{qubit}}$) is shown as a function of parameters governing
one-mode symplectic operators applied symmetrically.
That expressed directly by the hardware upon trapping without further operations,  $(z, \phi_1) = (1,0)$, is indicated in blue.
The configurations optimally localizing the domain of support to this qubit-sized subspace are indicated by solid and dashed black lines, achieved by locally transforming the covariance matrix to normal form with constant diagonal $\langle \hat{\ell}_j \hat{\ell}_j\rangle = \langle \hat{\pi}_j \hat{\pi}_j\rangle$.  (right) Scaling of probability outside the qudit subspace as a function of increasing Fock-space qudit dimension.
}
\label{fig:subspaceprobs2ions}
\end{figure}

While the presence of local operations does not impact the value of entanglement measures, they do impact the distribution of a CV quantum state in Fock space.
If aims of entanglement extraction from the local motional modes to the ion internal levels are pursued through exchange with the lowest Fock states, as first proposed by Ref.~\cite{Retzker2005:gor}, it is important to assure dominant support is present in this low-occupation Hilbert subspace.
For a two-ion chain, the left panel of Figure~\ref{fig:subspaceprobs2ions} demonstrates the deviation of this probability from unity, $P_{\overline{\text{qubit}}}$, i.e., the probability of the local motional modes being outside the two-state subspace of Fock states $|0\rangle$, and $|1\rangle$, as a function of local operations.
To parameterize single-mode symplectic operations,  one may perform a rudimentary application of the Bloch-Messiah decomposition---allowing a symplectic transformation to be decomposed into a set of single-mode squeezing~($Z$) operations surrounded by a pair of passive~($O$) operations (energy-preserving, e.g., beam-splitters and phase shifters)---with real numbers $(\phi_1, \phi_2, z)$ as~\cite{Arvind:1995symp,Braunstein:2005sqz,Serafini2017}
\begin{equation}
S = O_1ZO_2 \qquad , \qquad O_j = \begin{pmatrix}
\operatorname{Re}(\gamma_j) & \operatorname{Im}(\gamma_j) \\
-\operatorname{Im}(\gamma_j) & \operatorname{Re}(\gamma_j)
\end{pmatrix} \quad , \quad Z = \begin{pmatrix}
z & 0 \\
0 & \frac{1}{z}
\end{pmatrix} \ \ \ ,
\end{equation}
where $\gamma_j = e^{i \phi_j}$ is a one-dimensional unitary.
When squeezing each local motional mode by $z = \left(\sigma_{2,2}/\sigma_{1,1}\right)^{1/4}$, the quantum state upon trapping of an ion pair (blue point of Fig.~\ref{fig:subspaceprobs2ions}) can be shifted to a form that optimally compacts the support of the quantum state in the low Fock states (green point, solid line).
The alternate solution (dashed line) simply exchanges the position and momentum components, and thus both optimal points yield a two-mode squeezed vacuum state with Fock basis characterized by Eq.~\eqref{eq:tmsvsFockbasis}.
It is anticipated that this determination of the true form of the local motional mode CV quantum state, and its guidance of additional single-mode squeezing operations, will improve experimental implementations of ion motional entanglement extraction.

\section{Numerical Tables}
\label{app:numericalTables}
The following Tables~\ref{tab:1}-\ref{tab:end} provide numerical values presented in Figures~\ref{fig:logneg} and~\ref{fig:fidelities} of the main text and Figure~\ref{fig:subspaceprobs2ions} of Appendix~\ref{app:fockspace}.

\begin{table}[h]
\begin{tabular}{c|ccc|ccc}
\hline
\hline
  & \multicolumn{3}{c}{Ions} & \multicolumn{3}{c}{Scalar} \\
separation, $\tilde{r}/d$ & $\mathcal{N}_{A|B}\left(\sigma^{(t)}\right)$ & $\mathcal{N}_{A|B}\left(\sigma^{(m,\phi)}\right)$ & $\mathcal{N}_{A|B}\left(\sigma^{(m,\pi)}\right)$ & $\mathcal{N}_{A|B}\left(\sigma^{(t)}\right)$ & $\mathcal{N}_{A|B}\left(\sigma^{(m,\phi)}\right)$ & $\mathcal{N}_{A|B}\left(\sigma^{(m,\pi)}\right)$ \\
\hline
\hline
$0$ & $3.66\times 10^{-1}$ & $3.94\times 10^{-1}$ & $9.25\times 10^{-1}$ & $4.44\times 10^{-1}$ & $5.00\times 10^{-1}$ & $2.30 $ \\
$1$ & $5.95\times 10^{-2}$ & $1.14\times 10^{-1}$ & $7.38\times 10^{-1}$ & $0$ & $9.63\times 10^{-2}$ & $2.08 $ \\
$2$ & $0$ & $5.48\times 10^{-2}$ & $6.43\times 10^{-1}$ & $0$ & $4.12\times 10^{-2}$ & $1.97 $ \\
$3$ & $0$ & $3.25\times 10^{-2}$ & $5.81\times 10^{-1}$ & $0$ & $2.29\times 10^{-2}$ & $1.90 $ \\
$4$ & $0$ & $2.16\times 10^{-2}$ & $5.36\times 10^{-1}$ & $0$ & $1.46\times 10^{-2}$ & $1.85 $ \\
$5$ & $0$ & $1.54\times 10^{-2}$ & $5.00\times 10^{-1}$ & $0$ & $1.01\times 10^{-2}$ & $1.81 $ \\
$6$ & $0$ & $1.16\times 10^{-2}$ & $4.71\times 10^{-1}$ & $0$ & $7.40\times 10^{-3}$ & $1.78 $ \\
$7$ & $0$ & $9.04\times 10^{-3}$ & $4.46\times 10^{-1}$ & $0$ & $5.66\times 10^{-3}$ & $1.76 $ \\
$8$ & $0$ & $7.26\times 10^{-3}$ & $4.25\times 10^{-1}$ & $0$ & $4.47\times 10^{-3}$ & $1.73 $ \\
$9$ & $0$ & $5.97\times 10^{-3}$ & $4.06\times 10^{-1}$ & $0$ & $3.62\times 10^{-3}$ & $1.71 $ \\
$10$ & $0$ & $4.99\times 10^{-3}$ & $3.90\times 10^{-1}$ & $0$ & $2.99\times 10^{-3}$ & $1.70 $ \\
\hline
\hline
\end{tabular}
\caption{Numerical values for A, B region sizes of 1 ion or lattice site in Fig.~\ref{fig:logneg}.}
\label{tab:1}
\end{table}

\begin{table}
\begin{tabular}{c|ccc|ccc}
\hline
\hline
  & \multicolumn{3}{c}{Ions} & \multicolumn{3}{c}{Scalar} \\
separation, $\tilde{r}/d$ & $\mathcal{N}_{A|B}\left(\sigma^{(t)}\right)$ & $\mathcal{N}_{A|B}\left(\sigma^{(m,\phi)}\right)$ & $\mathcal{N}_{A|B}\left(\sigma^{(m,\pi)}\right)$ & $\mathcal{N}_{A|B}\left(\sigma^{(t)}\right)$ & $\mathcal{N}_{A|B}\left(\sigma^{(m,\phi)}\right)$ & $\mathcal{N}_{A|B}\left(\sigma^{(m,\pi)}\right)$ \\
\hline
\hline
$0$ & $6.29\times 10^{-1}$ & $6.73\times 10^{-1}$ & $1.22$ & $7.89\times 10^{-1}$ & $8.39\times 10^{-1}$ & $2.71 $ \\
$\frac{1}{3}$ & $2.01\times 10^{-1}$ & $2.47\times 10^{-1}$ & $9.91\times 10^{-1}$ & $1.68\times 10^{-1}$ & $2.19\times 10^{-1}$ & $2.43 $ \\
$\frac{2}{3}$ & $6.77\times 10^{-2}$ & $1.40\times 10^{-1}$ & $8.73\times 10^{-1}$ & $2.77\times 10^{-2}$ & $1.14\times 10^{-1}$ & $2.29 $ \\
$1$ & $1.27\times 10^{-2}$ & $9.25\times 10^{-2}$ & $7.95\times 10^{-1}$ & $4.41\times 10^{-3}$ & $7.21\times 10^{-2}$ & $2.20 $ \\
$\frac{4}{3}$ & $2.96\times 10^{-3}$ & $6.65\times 10^{-2}$ & $7.36\times 10^{-1}$ & $2.07\times 10^{-3}$ & $5.03\times 10^{-2}$ & $2.14 $ \\
$\frac{5}{3}$ & $1.45\times 10^{-3}$ & $5.05\times 10^{-2}$ & $6.90\times 10^{-1}$ & $1.19\times 10^{-3}$ & $3.73\times 10^{-2}$ & $2.09 $ \\
$2$ & $8.58\times 10^{-4}$ & $3.98\times 10^{-2}$ & $6.52\times 10^{-1}$ & $6.99\times 10^{-4}$ & $2.88\times 10^{-2}$ & $2.05 $ \\
$\frac{7}{3}$ & $5.31\times 10^{-4}$ & $3.23\times 10^{-2}$ & $6.19\times 10^{-1}$ & $3.86\times 10^{-4}$ & $2.30\times 10^{-2}$ & $2.02 $ \\
$\frac{8}{3}$ & $3.20\times 10^{-4}$ & $2.68\times 10^{-2}$ & $5.91\times 10^{-1}$ & $1.65\times 10^{-4}$ & $1.88\times 10^{-2}$ & $1.99 $ \\
$3$ & $1.71\times 10^{-4}$ & $2.26\times 10^{-2}$ & $5.66\times 10^{-1}$ & $0$ & $1.57\times 10^{-2}$ & $1.96 $ \\
$\frac{10}{3}$ & $5.99\times 10^{-5}$ & $1.93\times 10^{-2}$ & $5.44\times 10^{-1}$ & $0$ & $1.33\times 10^{-2}$ & $1.94 $ \\
$\frac{11}{3}$ & $0$ & $1.67\times 10^{-2}$ & $5.24\times 10^{-1}$ & $0$ & $1.14\times 10^{-2}$ & $1.92 $ \\
$4$ & $0$ & $1.46\times 10^{-2}$ & $5.06\times 10^{-1}$ & $0$ & $9.88\times 10^{-3}$ & $1.90 $ \\
$\frac{13}{3}$ & $0$ & $1.29\times 10^{-2}$ & $4.89\times 10^{-1}$ & $0$ & $8.65\times 10^{-3}$ & $1.89 $ \\
$\frac{14}{3}$ & $0$ & $1.15\times 10^{-2}$ & $4.74\times 10^{-1}$ & $0$ & $7.64\times 10^{-3}$ & $1.87 $ \\
$5$ & $0$ & $1.03\times 10^{-2}$ & $4.59\times 10^{-1}$ & $0$ & $6.80\times 10^{-3}$ & $1.86 $ \\
$\frac{16}{3}$ & $0$ & $9.28\times 10^{-3}$ & $4.46\times 10^{-1}$ & $0$ & $6.09\times 10^{-3}$ & $1.85 $ \\
$\frac{17}{3}$ & $0$ & $8.41\times 10^{-3}$ & $4.33\times 10^{-1}$ & $0$ & $5.49\times 10^{-3}$ & $1.83 $ \\
$6$ & $0$ & $7.66\times 10^{-3}$ & $4.22\times 10^{-1}$ & $0$ & $4.97\times 10^{-3}$ & $1.82 $ \\
$\frac{19}{3}$ & $0$ & $7.01\times 10^{-3}$ & $4.10\times 10^{-1}$ & $0$ & $4.53\times 10^{-3}$ & $1.81 $ \\
$\frac{20}{3}$ & $0$ & $6.43\times 10^{-3}$ & $4.00\times 10^{-1}$ & $0$ & $4.14\times 10^{-3}$ & $1.80 $ \\
$7$ & $0$ & $5.93\times 10^{-3}$ & $3.90\times 10^{-1}$ & $0$ & $3.80\times 10^{-3}$ & $1.79 $ \\
$\frac{22}{3}$ & $0$ & $5.48\times 10^{-3}$ & $3.81\times 10^{-1}$ & $0$ & $3.50\times 10^{-3}$ & $1.78 $ \\
$\frac{23}{3}$ & $0$ & $5.09\times 10^{-3}$ & $3.72\times 10^{-1}$ & $0$ & $3.23\times 10^{-3}$ & $1.78 $ \\
$8$ & $0$ & $4.73\times 10^{-3}$ & $3.63\times 10^{-1}$ & $0$ & $2.99\times 10^{-3}$ & $1.77 $ \\
$\frac{25}{3}$ & $0$ & $4.41\times 10^{-3}$ & $3.55\times 10^{-1}$ & $0$ & $2.78\times 10^{-3}$ & $1.76 $ \\
$\frac{26}{3}$ & $0$ & $4.13\times 10^{-3}$ & $3.47\times 10^{-1}$ & $0$ & $2.59\times 10^{-3}$ & $1.75 $ \\
$9$ & $0$ & $3.87\times 10^{-3}$ & $3.39\times 10^{-1}$ & $0$ & $2.42\times 10^{-3}$ & $1.75 $ \\
\hline
\hline
\end{tabular}
\caption{Numerical values for A, B region sizes of 3 ions or lattice sites in Fig.~\ref{fig:logneg}.}
\end{table}

\begin{table}
\begin{tabular}{c|ccc|ccc}
\hline
\hline
  & \multicolumn{3}{c}{Ions} & \multicolumn{3}{c}{Scalar} \\
separation, $\tilde{r}/d$ & $\mathcal{N}_{A|B}\left(\sigma^{(t)}\right)$ & $\mathcal{N}_{A|B}\left(\sigma^{(m,\phi)}\right)$ & $\mathcal{N}_{A|B}\left(\sigma^{(m,\pi)}\right)$ & $\mathcal{N}_{A|B}\left(\sigma^{(t)}\right)$ & $\mathcal{N}_{A|B}\left(\sigma^{(m,\phi)}\right)$ & $\mathcal{N}_{A|B}\left(\sigma^{(m,\pi)}\right)$ \\
\hline
\hline
$0$ & $7.75\times 10^{-1}$ & $8.18\times 10^{-1}$ & $1.33$ & $9.67\times 10^{-1}$ & $1.01$ & $2.89 $ \\
$\frac{1}{5}$ & $2.89\times 10^{-1}$ & $3.32\times 10^{-1}$ & $1.09$ & $2.55\times 10^{-1}$ & $3.00\times 10^{-1}$ & $2.59 $ \\
$\frac{2}{5}$ & $1.32\times 10^{-1}$ & $2.02\times 10^{-1}$ & $9.69\times 10^{-1}$ & $8.55\times 10^{-2}$ & $1.71\times 10^{-1}$ & $2.44 $ \\
$\frac{3}{5}$ & $5.77\times 10^{-2}$ & $1.41\times 10^{-1}$ & $8.85\times 10^{-1}$ & $2.87\times 10^{-2}$ & $1.15\times 10^{-1}$ & $2.35 $ \\
$\frac{4}{5}$ & $2.36\times 10^{-2}$ & $1.06\times 10^{-1}$ & $8.22\times 10^{-1}$ & $1.40\times 10^{-2}$ & $8.42\times 10^{-2}$ & $2.28 $ \\
$1$ & $1.17\times 10^{-2}$ & $8.32\times 10^{-2}$ & $7.72\times 10^{-1}$ & $8.53\times 10^{-3}$ & $6.48\times 10^{-2}$ & $2.22 $ \\
$\frac{6}{5}$ & $7.11\times 10^{-3}$ & $6.74\times 10^{-2}$ & $7.30\times 10^{-1}$ & $5.66\times 10^{-3}$ & $5.17\times 10^{-2}$ & $2.18 $ \\
$\frac{7}{5}$ & $4.74\times 10^{-3}$ & $5.60\times 10^{-2}$ & $6.95\times 10^{-1}$ & $3.86\times 10^{-3}$ & $4.24\times 10^{-2}$ & $2.14 $ \\
$\frac{8}{5}$ & $3.30\times 10^{-3}$ & $4.73\times 10^{-2}$ & $6.64\times 10^{-1}$ & $2.63\times 10^{-3}$ & $3.54\times 10^{-2}$ & $2.11 $ \\
$\frac{9}{5}$ & $2.32\times 10^{-3}$ & $4.06\times 10^{-2}$ & $6.37\times 10^{-1}$ & $1.73\times 10^{-3}$ & $3.01\times 10^{-2}$ & $2.08 $ \\
$2$ & $1.62\times 10^{-3}$ & $3.53\times 10^{-2}$ & $6.13\times 10^{-1}$ & $1.06\times 10^{-3}$ & $2.59\times 10^{-2}$ & $2.06 $ \\
$\frac{11}{5}$ & $1.09\times 10^{-3}$ & $3.10\times 10^{-2}$ & $5.91\times 10^{-1}$ & $5.78\times 10^{-4}$ & $2.26\times 10^{-2}$ & $2.04 $ \\
$\frac{12}{5}$ & $6.87\times 10^{-4}$ & $2.74\times 10^{-2}$ & $5.70\times 10^{-1}$ & $2.65\times 10^{-4}$ & $1.98\times 10^{-2}$ & $2.02 $ \\
$\frac{13}{5}$ & $3.86\times 10^{-4}$ & $2.45\times 10^{-2}$ & $5.52\times 10^{-1}$ & $9.64\times 10^{-5}$ & $1.76\times 10^{-2}$ & $2.00 $ \\
$\frac{14}{5}$ & $1.83\times 10^{-4}$ & $2.20\times 10^{-2}$ & $5.35\times 10^{-1}$ & $2.62\times 10^{-5}$ & $1.57\times 10^{-2}$ & $1.98 $ \\
$3$ & $6.87\times 10^{-5}$ & $1.99\times 10^{-2}$ & $5.19\times 10^{-1}$ & $9.51\times 10^{-6}$ & $1.41\times 10^{-2}$ & $1.97 $ \\
$\frac{16}{5}$ & $2.00\times 10^{-5}$ & $1.80\times 10^{-2}$ & $5.04\times 10^{-1}$ & $5.13\times 10^{-6}$ & $1.27\times 10^{-2}$ & $1.95 $ \\
$\frac{17}{5}$ & $7.42\times 10^{-6}$ & $1.65\times 10^{-2}$ & $4.90\times 10^{-1}$ & $3.28\times 10^{-6}$ & $1.16\times 10^{-2}$ & $1.94 $ \\
$\frac{18}{5}$ & $4.01\times 10^{-6}$ & $1.51\times 10^{-2}$ & $4.77\times 10^{-1}$ & $2.27\times 10^{-6}$ & $1.05\times 10^{-2}$ & $1.93 $ \\
$\frac{19}{5}$ & $2.59\times 10^{-6}$ & $1.39\times 10^{-2}$ & $4.65\times 10^{-1}$ & $1.63\times 10^{-6}$ & $9.66\times 10^{-3}$ & $1.91 $ \\
$4$ & $1.82\times 10^{-6}$ & $1.28\times 10^{-2}$ & $4.53\times 10^{-1}$ & $1.18\times 10^{-6}$ & $8.88\times 10^{-3}$ & $1.90 $ \\
$\frac{21}{5}$ & $1.33\times 10^{-6}$ & $1.19\times 10^{-2}$ & $4.42\times 10^{-1}$ & $8.56\times 10^{-7}$ & $8.20\times 10^{-3}$ & $1.89 $ \\
$\frac{22}{5}$ & $9.99\times 10^{-7}$ & $1.10\times 10^{-2}$ & $4.32\times 10^{-1}$ & $6.05\times 10^{-7}$ & $7.59\times 10^{-3}$ & $1.88 $ \\
$\frac{23}{5}$ & $7.55\times 10^{-7}$ & $1.03\times 10^{-2}$ & $4.21\times 10^{-1}$ & $4.05\times 10^{-7}$ & $7.05\times 10^{-3}$ & $1.87 $ \\
$\frac{24}{5}$ & $5.69\times 10^{-7}$ & $9.61\times 10^{-3}$ & $4.12\times 10^{-1}$ & $2.41\times 10^{-7}$ & $6.56\times 10^{-3}$ & $1.87 $ \\
$5$ & $4.21\times 10^{-7}$ & $9.00\times 10^{-3}$ & $4.03\times 10^{-1}$ & $1.05\times 10^{-7}$ & $6.12\times 10^{-3}$ & $1.86 $ \\
$\frac{26}{5}$ & $3.01\times 10^{-7}$ & $8.45\times 10^{-3}$ & $3.94\times 10^{-1}$ & $0$ & $5.73\times 10^{-3}$ & $1.85 $ \\
$\frac{27}{5}$ & $2.01\times 10^{-7}$ & $7.94\times 10^{-3}$ & $3.85\times 10^{-1}$ & $0$ & $5.37\times 10^{-3}$ & $1.84 $ \\
$\frac{28}{5}$ & $1.17\times 10^{-7}$ & $7.48\times 10^{-3}$ & $3.77\times 10^{-1}$ & $0$ & $5.04\times 10^{-3}$ & $1.83 $ \\
$\frac{29}{5}$ & $4.50\times 10^{-8}$ & $7.07\times 10^{-3}$ & $3.69\times 10^{-1}$ & $0$ & $4.75\times 10^{-3}$ & $1.83 $ \\
$6$ & $0$ & $6.68\times 10^{-3}$ & $3.62\times 10^{-1}$ & $0$ & $4.48\times 10^{-3}$ & $1.82 $ \\
$\frac{31}{5}$ & $0$ & $6.33\times 10^{-3}$ & $3.55\times 10^{-1}$ & $0$ & $4.23\times 10^{-3}$ & $1.81 $ \\
$\frac{32}{5}$ & $0$ & $6.00\times 10^{-3}$ & $3.48\times 10^{-1}$ & $0$ & $4.00\times 10^{-3}$ & $1.81 $ \\
$\frac{33}{5}$ & $0$ & $5.70\times 10^{-3}$ & $3.41\times 10^{-1}$ & $0$ & $3.79\times 10^{-3}$ & $1.80 $ \\
$\frac{34}{5}$ & $0$ & $5.43\times 10^{-3}$ & $3.34\times 10^{-1}$ & $0$ & $3.59\times 10^{-3}$ & $1.80 $ \\
$7$ & $0$ & $5.17\times 10^{-3}$ & $3.28\times 10^{-1}$ & $0$ & $3.42\times 10^{-3}$ & $1.79 $ \\
$\frac{36}{5}$ & $0$ & $4.93\times 10^{-3}$ & $3.22\times 10^{-1}$ & $0$ & $3.25\times 10^{-3}$ & $1.78 $ \\
$\frac{37}{5}$ & $0$ & $4.71\times 10^{-3}$ & $3.16\times 10^{-1}$ & $0$ & $3.10\times 10^{-3}$ & $1.78 $ \\
$\frac{38}{5}$ & $0$ & $4.50\times 10^{-3}$ & $3.10\times 10^{-1}$ & $0$ & $2.95\times 10^{-3}$ & $1.77 $ \\
$\frac{39}{5}$ & $0$ & $4.30\times 10^{-3}$ & $3.04\times 10^{-1}$ & $0$ & $2.82\times 10^{-3}$ & $1.77 $ \\
$8$ & $0$ & $4.12\times 10^{-3}$ & $2.99\times 10^{-1}$ & $0$ & $2.69\times 10^{-3}$ & $1.76 $ \\
$\frac{41}{5}$ & $0$ & $3.95\times 10^{-3}$ & $2.93\times 10^{-1}$ & $0$ & $2.58\times 10^{-3}$ & $1.76 $ \\
$\frac{42}{5}$ & $0$ & $3.79\times 10^{-3}$ & $2.88\times 10^{-1}$ & $0$ & $2.47\times 10^{-3}$ & $1.75 $ \\
$\frac{43}{5}$ & $0$ & $3.65\times 10^{-3}$ & $2.83\times 10^{-1}$ & $0$ & $2.36\times 10^{-3}$ & $1.75 $ \\
$\frac{44}{5}$ & $0$ & $3.50\times 10^{-3}$ & $2.78\times 10^{-1}$ & $0$ & $2.27\times 10^{-3}$ & $1.75 $ \\
$9$ & $0$ & $3.37\times 10^{-3}$ & $2.73\times 10^{-1}$ & $0$ & $2.18\times 10^{-3}$ & $1.74 $ \\
\hline
\hline
\end{tabular}
\caption{Numerical values for A, B region sizes of 5 ions or lattice sites in Fig.~\ref{fig:logneg}.}
\end{table}

\begin{table}
\begin{tabular}{c|c|c|c}
\hline
\hline
& & \multicolumn{2}{c}{Fidelity}\\
Region Size & Squeezing, $z$ &  Raw &  Locally Squeezed \\
\hline
\hline
$2$ & $4.119$ & $3.49\times 10^{-1}$ & $0.814 $ \\
$4$ & $3.666$ & $1.88\times 10^{-1}$ & $0.767 $ \\
$6$ & $3.507$ & $1.03\times 10^{-1}$ & $0.738 $ \\
$8$ & $3.420$ & $5.79\times 10^{-2}$ & $0.717 $ \\
$10$ & $3.362$ & $3.28\times 10^{-2}$ & $0.699 $ \\
$12$ & $3.315$ & $1.89\times 10^{-2}$ & $0.682 $ \\
$14$ & $3.274$ & $1.10\times 10^{-2}$ & $0.667 $ \\
$16$ & $3.234$ & $6.56\times 10^{-3}$ & $0.652 $ \\
$18$ & $3.194$ & $3.99\times 10^{-3}$ & $0.636 $ \\
$20$ & $3.152$ & $2.49\times 10^{-3}$ & $0.619 $ \\
$22$ & $3.106$ & $1.60\times 10^{-3}$ & $0.600 $ \\
$24$ & $3.054$ & $1.08\times 10^{-3}$ & $0.576 $ \\
$26$ & $2.993$ & $7.59\times 10^{-4}$ & $0.544 $ \\
$28$ & $2.919$ & $5.73\times 10^{-4}$ & $0.493 $ \\
$30$ & $2.821$ & $4.47\times 10^{-4}$ & $0.368 $ \\
\hline
\hline
\end{tabular}
\caption{Numerical values for the $N = 30$ ion chain shown in Fig.~\ref{fig:fidelities}.}
\end{table}

\begin{table}
\begin{tabular}{c|c|c|c}
\hline
\hline
& & \multicolumn{2}{c}{Fidelity}\\
Region Size & Squeezing, $z$ & Raw &  Locally Squeezed \\
\hline
\hline
$2$ & $5.118$ & $2.83\times 10^{-1}$ & $0.827 $ \\
$4$ & $4.579$ & $1.25\times 10^{-1}$ & $0.784 $ \\
$6$ & $4.390$ & $5.70\times 10^{-2}$ & $0.758 $ \\
$8$ & $4.292$ & $2.62\times 10^{-2}$ & $0.739 $ \\
$10$ & $4.229$ & $1.22\times 10^{-2}$ & $0.724 $ \\
$12$ & $4.184$ & $5.67\times 10^{-3}$ & $0.710 $ \\
$14$ & $4.149$ & $2.67\times 10^{-3}$ & $0.698 $ \\
$16$ & $4.118$ & $1.26\times 10^{-3}$ & $0.686 $ \\
$18$ & $4.091$ & $6.04\times 10^{-4}$ & $0.675 $ \\
$20$ & $4.066$ & $2.91\times 10^{-4}$ & $0.665 $ \\
$22$ & $4.041$ & $1.42\times 10^{-4}$ & $0.654 $ \\
$24$ & $4.016$ & $6.97\times 10^{-5}$ & $0.644 $ \\
$26$ & $3.991$ & $3.47\times 10^{-5}$ & $0.634 $ \\
$28$ & $3.965$ & $1.75\times 10^{-5}$ & $0.624 $ \\
$30$ & $3.938$ & $9.00\times 10^{-6}$ & $0.613 $ \\
$32$ & $3.910$ & $4.70\times 10^{-6}$ & $0.601 $ \\
$34$ & $3.879$ & $2.51\times 10^{-6}$ & $0.588 $ \\
$36$ & $3.846$ & $1.37\times 10^{-6}$ & $0.574 $ \\
$38$ & $3.811$ & $7.67\times 10^{-7}$ & $0.557 $ \\
$40$ & $3.771$ & $4.44\times 10^{-7}$ & $0.538 $ \\
$42$ & $3.728$ & $2.66\times 10^{-7}$ & $0.515 $ \\
$44$ & $3.679$ & $1.67\times 10^{-7}$ & $0.485 $ \\
$46$ & $3.623$ & $1.12\times 10^{-7}$ & $0.445 $ \\
$48$ & $3.556$ & $8.03\times 10^{-8}$ & $0.387 $ \\
$50$ & $3.473$ & $6.09\times 10^{-8}$ & $0.271 $ \\
\hline
\hline
\end{tabular}
\caption{Numerical values for the $N = 50$ ion chain shown in Fig.~\ref{fig:fidelities}.}
\end{table}

\begin{table}
\begin{tabular}{c|c|c|c}
\hline
\hline
& & \multicolumn{2}{c}{Fidelity}\\
Region Size & Squeezing, $z$ &  Raw &  Locally Squeezed \\
\hline
\hline
$2$ & $8.262$ & $1.73\times 10^{-1}$ & $0.850 $ \\
$4$ & $7.453$ & $4.86\times 10^{-2}$ & $0.813 $ \\
$6$ & $7.166$ & $1.40\times 10^{-2}$ & $0.792 $ \\
$8$ & $7.018$ & $4.08\times 10^{-3}$ & $0.776 $ \\
$10$ & $6.928$ & $1.19\times 10^{-3}$ & $0.763 $ \\
$12$ & $6.866$ & $3.51\times 10^{-4}$ & $0.752 $ \\
$14$ & $6.821$ & $1.03\times 10^{-4}$ & $0.742 $ \\
$16$ & $6.787$ & $3.06\times 10^{-5}$ & $0.733 $ \\
$18$ & $6.759$ & $9.05\times 10^{-6}$ & $0.724 $ \\
$20$ & $6.736$ & $2.68\times 10^{-6}$ & $0.716 $ \\
$22$ & $6.716$ & $7.98\times 10^{-7}$ & $0.709 $ \\
$24$ & $6.699$ & $2.37\times 10^{-7}$ & $0.701 $ \\
$26$ & $6.684$ & $7.08\times 10^{-8}$ & $0.694 $ \\
$28$ & $6.671$ & $2.12\times 10^{-8}$ & $0.688 $ \\
$30$ & $6.658$ & $6.34\times 10^{-9}$ & $0.681 $ \\
$32$ & $6.646$ & $1.90\times 10^{-9}$ & $0.675 $ \\
$34$ & $6.635$ & $5.71\times 10^{-10}$ & $0.669 $ \\
$36$ & $6.625$ & $1.72\times 10^{-10}$ & $0.663 $ \\
$38$ & $6.615$ & $5.19\times 10^{-11}$ & $0.657 $ \\
$40$ & $6.605$ & $1.57\times 10^{-11}$ & $0.652 $ \\
$42$ & $6.596$ & $4.76\times 10^{-12}$ & $0.646 $ \\
$44$ & $6.586$ & $1.45\times 10^{-12}$ & $0.640 $ \\
$46$ & $6.577$ & $4.41\times 10^{-13}$ & $0.635 $ \\
$48$ & $6.568$ & $1.35\times 10^{-13}$ & $0.630 $ \\
$50$ & $6.558$ & $4.12\times 10^{-14}$ & $0.624 $ \\
\hline
\hline
\end{tabular}
\caption{Numerical values for the $N = 150$ ion chain shown in Fig.~\ref{fig:fidelities}.}
\end{table}

\begin{table}
\begin{tabular}{c|c|c}
\hline
\hline
 & \multicolumn{2}{c}{ $P_{\overline{qudit}}$}
\\
Fock truncation/qudit dimension& Raw &  Locally Squeezed \\
\hline
\hline
$2$ & $1.93\times 10^{-2}$ & $3.47\times 10^{-4} $ \\
$3$ & $1.28\times 10^{-3}$ & $6.46\times 10^{-6} $ \\
$4$ & $3.17\times 10^{-4}$ & $1.20\times 10^{-7} $ \\
$5$ & $2.99\times 10^{-5}$ & $2.24\times 10^{-9} $ \\
$6$ & $5.88\times 10^{-6}$ & $4.17\times 10^{-11} $ \\
$7$ & $6.73\times 10^{-7}$ & $7.77\times 10^{-13} $ \\
$8$ & $1.17\times 10^{-7}$ & $1.45\times 10^{-14} $ \\
\hline
\hline
\end{tabular}
\caption{Numerical values shown in Fig.~\ref{fig:subspaceprobs2ions}.}
\label{tab:end}
\end{table}

\end{document}